\documentclass[amsmath,amssymb,superscriptaddress,nobalancelastpage,prl,twocolumn,preprintnumbers, numbers]{revtex4-1}

\usepackage{bm}
\usepackage{braket}
\usepackage{diagbox}
\usepackage{graphicx}
\usepackage{booktabs}
\usepackage{hyperref}
\usepackage{varioref}
\usepackage{xr-hyper}
\hypersetup{colorlinks,linkcolor=blue,urlcolor=blue,citecolor=blue}
\usepackage{siunitx}
\usepackage{ulem}
\usepackage{alphabeta}
\usepackage{orcidlink}
\usepackage{xcolor}

\newcommand{\green}[1]{{\textcolor[rgb]{0,0.5,0}{#1}}}

\newcommand{\EuAl}{EuAl$_4$}

\begin{document}

\author{M. Baumgartner~\orcidlink{0009-0003-9070-6639}}
\thanks{These authors contributed equally.}
\affiliation{Physik-Institut, Universit\"{a}t Z\"{u}rich, Winterthurerstrasse 190, CH-8057 Z\"{u}rich, Switzerland}

\author{Xunyang~Hong~\orcidlink{0000-0001-6219-2851}}
\thanks{These authors contributed equally.}
\affiliation{Department of Physics, The Chinese University of Hong Kong, Shatin, Hong Kong, China}
\affiliation{Physik-Institut, Universit\"{a}t Z\"{u}rich, Winterthurerstrasse 190, CH-8057 Z\"{u}rich, Switzerland}

\author{Tianren Wang}
\thanks{These authors contributed equally.}
\affiliation{Department of Physics, The Chinese University of Hong Kong, Shatin, Hong Kong, China}
\affiliation{State Key Laboratory of Quantum Information Technologies and Materials, The Chinese University of Hong Kong, Shatin, Hong Kong, China}

\author{Fazhi Yang}
\affiliation{Department of Physics, City University of Hong Kong, Kowloon, Hong Kong, China}

\author{Yuetong Wu}
\affiliation{Department of Physics, The Chinese University of Hong Kong, Shatin, Hong Kong, China}
\affiliation{State Key Laboratory of  Quantum Information Technologies and Materials, The Chinese University of Hong Kong, Shatin, Hong Kong, China}

\author{Junzhang Ma}
\affiliation{Department of Physics, City University of Hong Kong, Kowloon, Hong Kong, China}

\author{Tian Shang}
\affiliation{School of Physics, East China Normal University, Shanghai 200241, China}

\author{J.~Oppliger}
\affiliation{Physik-Institut, Universit\"{a}t Z\"{u}rich, Winterthurerstrasse 190, CH-8057 Z\"{u}rich, Switzerland}

\author{J.~Küspert}
\affiliation{Physik-Institut, Universit\"{a}t Z\"{u}rich, Winterthurerstrasse 190, CH-8057 Z\"{u}rich, Switzerland}

\author{M.~Hücker}
\affiliation{Department of Condensed Matter Physics, Weizmann Institute of Science, Rehovot, 7610001, Israel.}

\author{O.~Ivashko}
\affiliation{Physik-Institut, Universit\"{a}t Z\"{u}rich, Winterthurerstrasse 190, CH-8057 Z\"{u}rich, Switzerland}

\author{F.~Igoa Saldaña}
\affiliation{Deutsches Elektronen-Synchrotron DESY, Notkestra{\ss}e 85, 22607 Hamburg, Germany}

\author{M. v. Zimmermann}
\affiliation{Deutsches Elektronen-Synchrotron DESY, Notkestra{\ss}e 85, 22607 Hamburg, Germany}

\author{S.~Pyon~\orcidlink{0000-0002-5716-1791}}
\affiliation{Department of Applied Physics, The University of Tokyo, Tokyo, 113-8656, Japan}

\author{K.~Kudo~\orcidlink{0000-0003-2839-6791}}
\affiliation{Department of Physics,
The University of Osaka,
1-1 Machikaneyama, Toyonaka, Osaka 560-0043, Japan}
\affiliation{Institute for Open and Transdisciplinary Research Initiatives,
The University of Osaka,
2-1 Yamadaoka, Suita, Osaka 565-0871, Japan}

\author{M.~Nohara~\orcidlink{0000-0003-1931-9903}}
\affiliation{Department of Quantum Matter, Hiroshima University, Higashi-Hiroshima 739-8530, Japan}

\author{P. Sa\v{c}er}
\affiliation{Department of Physics, Faculty of Science, University of Zagreb, Bijeni\v{c}ka 32, HR-10000 Zagreb, Croatia}

\author{A. Akrap}
\affiliation{Department of Physics, Faculty of Science, University of Zagreb, Bijeni\v{c}ka 32, HR-10000 Zagreb, Croatia}

\author{N. Bari\v{s}i\'{c}}
\affiliation{Department of Physics, Faculty of Science, University of Zagreb, Bijeni\v{c}ka 32, HR-10000 Zagreb, Croatia}
\affiliation{Institute of Solid State Physics, TU Wien, A-1040 Vienna, Austria}

\author{M. Novak}
\affiliation{Department of Physics, Faculty of Science, University of Zagreb, Bijeni\v{c}ka 32, HR-10000 Zagreb, Croatia}

\author{Qisi Wang~\orcidlink{0000-0002-8741-7559}}
\thanks{Corresponding author: Qisi Wang}
\email{qwang@cuhk.edu.hk}
\affiliation{Department of Physics, The Chinese University of Hong Kong, Shatin, Hong Kong, China}
\affiliation{State Key Laboratory of Quantum Information Technologies and Materials, The Chinese University of Hong Kong, Shatin, Hong Kong, China}
   
\author{J. Chang~\orcidlink{0000-0002-4655-1516}}
\affiliation{Physik-Institut, Universit\"{a}t Z\"{u}rich, Winterthurerstrasse 190, CH-8057 Z\"{u}rich, Switzerland}

\title{Strain-tunable charge localization coupled to complex magnetic orders in EuAl$_4$}

\maketitle
\textbf{Charge localization is particularly 
interesting when 
coupled to antiferromagnetic spin structures. Coupled spin-charge 
orders are well established in elemental chromium and correlated oxide superconductors, yet the interplay between charge order and more complex magnetic textures — such as skyrmion lattices and chiral spin structures — remains largely unexplored. Here we report a 
comprehensive study of how charge localization couples to the unusually rich sequence of magnetic phases in EuAl$_4$. Using x-ray diffraction under applied magnetic field and uniaxial pressure, we demonstrate a direct coupling between the charge and spin order parameters. In the absence of external 
stimuli, charge localization is markedly enhanced upon entering the magnetically ordered phases. Strikingly, this effect is highly susceptible to strain: 
uniaxial pressure applied along the charge-order propagation direction further enhances localization, whereas pressure applied perpendicular to it weakens it. 
Application of magnetic field reveals both competitive and possible collaborative interactions between spin and charge ordering.
This flexible coupling between spin and charge ordering opens a new route  to designing  symmetry-breaking states. Chiral charge order may for example be patterned from spin structures with that symmetry.}

\textit{Introduction:} Charge ordering is a broad term encompassing a variety of lattice symmetry-breaking phenomena involving the spatial redistribution of electronic charge. At one extreme, charge density wave (CDW) instabilities in metals are well understood as weak-coupling phenomena driven by Fermi surface nesting, in which a divergence in the electronic susceptibility at a particular wavevector $\mathbf{q}$
 drives a sinusoidal modulation of the charge density with a periodicity incommensurate with the underlying lattice~\cite{gruner_dynamics_1988,monceau_electronic_2012}. A canonical example is the transition metal dichalcogenide NbSe$_2$, which below $T\mathrm{_{CDW}}=33$~K
 develops such a modulation while simultaneously entering a superconducting state at $T_c=7.2$~K, making it a paradigmatic system for studies of co-existing  
 orders~\cite{kiss2007Chargeordermaximized,xi2015Strongly,kundu2024Charge}.

\begin{figure*}
    \centering
    \includegraphics[width=\textwidth]{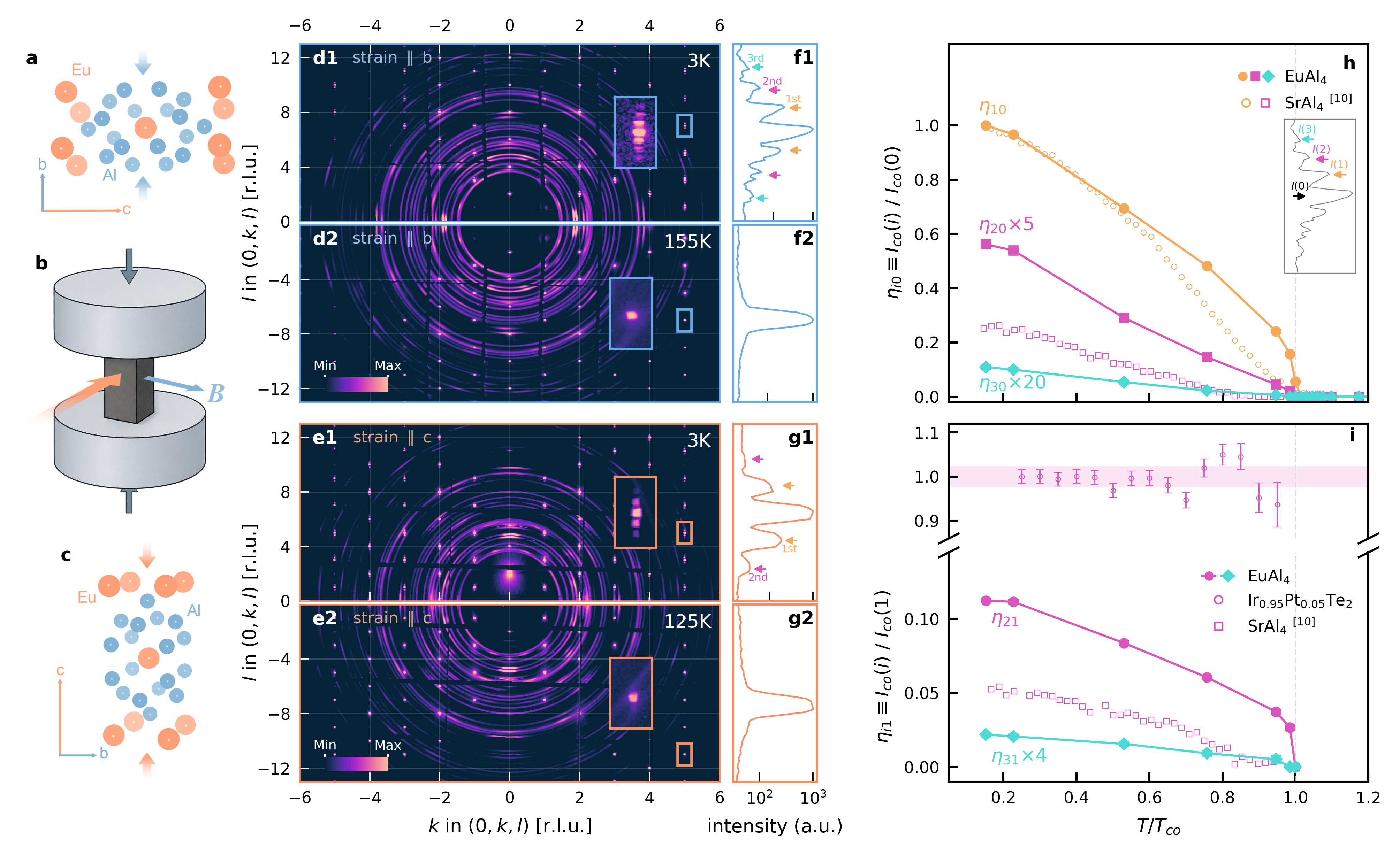}
    \caption{\textbf{Charge ordering in \EuAl.}
    (a, c) Sketch of the EuAl$_4$ crystal structure and the vertical applied uniaxial pressure from the strain cell shown in (b). Orange arrow depicts the in-coming beam, with horizontal magnetic field $\mathbf{B}$ (blue arrow) perpendicular to it.  
    (d1, d2, e1, e2) Reconstructed X-ray diffraction intensity within the $(0,k,\ell)$  plane at indicated temperatures, with crystal orientation and applied pressure depicted in (a, c) respectively. The insets display the raw diffraction data zoomed in around fundamental Bragg reflections at indicated regions. Charge order manifests through incommensurate reflections along the reciprocal $c$-axis direction.
    (f1, f2, g1, g2) Line profiles through the highlighted regions in log scale, showing the first, second, and third harmonics of the charge order satellite peaks, as marked by arrows. (h) Temperature dependence of the normalized first-, second-, and third-order satellite intensities, $\eta_{i0}\equiv I_{co}(i)/I_{co}(0)$,  $i=1, 2,3$, compared with results obtained on SrAl$_4$ from Ref. \cite{saraf2025Tuning}. The inset shows the definition of the intensity $I_{co}(i)$.
    (i) The intensity ratio $\eta_{21}$ and $\eta_{31}$ as functions of the reduced temperature $T/T_{co}$, compared with results  obtained on SrAl$_4$ (Ref.~\cite{saraf2025Tuning}) and Ir$_{0.95}$Pt$_{0.05}$Te$_2$.
    Horizontal purple line is a guide to the eye. Error bars, representing the fitting uncertainty, are smaller than the symbol size for \EuAl.}
    \label{figure1}
\end{figure*}
 
At the opposite extreme lies a strongly localized charge order, in which electrons form dimers~\cite{katukuri2022Charge, hwang2022Largegap}, trimers~\cite{senn_charge_2012}, or more complex real-space patterns~\cite{radaelli2002Formation}. The localization in such cases can be so pronounced that individual charge-ordered domains behave as coherent scatterers, enabling Young's double-slit-type interference experiments that directly visualize the ordered superstructure~\cite{revelli2019Resonant, magnaterra2023RIXS}. This regime is often realized in compounds containing heavy elements with strong spin-orbit coupling, a prominent example being IrTe$_2$, where stripe-like dimerization of Ir–Ir bonds produces a cascade of first-order phase transitions upon cooling~\cite{pascut2014DimerizationInduced,ivashko2017ChargeStripe}.

Diffraction experiments provide a natural way to distinguish between these two limiting cases. A conventional density wave, being a sinusoidal modulation of charge density in real space, generates only a single pair of satellite reflections at $\pm\mathbf{q}$ relative to a structural Bragg peak, with no higher harmonics. Strongly localized charge order, on the contrary, is characterized by a non-sinusoidal real-space profile, which manifests in reciprocal space as a series of intense higher harmonics at $\pm n\mathbf{q}$ ($n=2,3\ldots$). The relative intensities of these harmonics thus serve as a quantitative measure of the degree of charge localization, interpolating between the weak-coupling CDW limit and the strongly correlated, site-localized limit.

Between these two extremes lies a rich landscape of intermediate and intertwined ordering phenomena. 
Elementary chromium is an archetypal example: below its Néel temperature ($T_N = 311$~K), it develops a spin density wave (SDW), which is accompanied by a CDW at twice the SDW wavevector~\cite{jaramillo_breakdown_2009a} --- a direct consequence of the spin-charge coupling 
~\cite{fawcett1988Spindensitywave}. 
A more complex example is the correlated kagome metal FeGe, where CDW is found to couple to antiferromagnetic order, suggesting an intriguing interplay between magnetism and electronic bands~\cite{teng2022Discovery}. 
Coexisting spin and charge stripe order has also been identified in oxide compounds such as La$_{2-x}$Sr$_{x}$NiO$_4$~\cite{yoshizawa_stripe_2000} and  La$_{1.875}$Ba$_{0.125}$CuO$_4$~\cite{tranquada_evidence_1995,abbamonte2005Spatially, fink_phase_2011,vonarx_fate_2023}.
The charge modulation in these stripes appears to be more strongly localized than what would be expected for a simple density wave. Indeed, higher harmonics reflections of charge order are observed in the nickelates systems~\cite{yoshizawa_stripe_2000}. However, in the cuprate system, they have not yet been reported to date, leaving the true character of the charge modulation an open question. In this context, materials in which charge order couples to magnetic  structures offer a particularly valuable opportunity to investigate  the mechanisms driving charge localization and its response to external tuning parameters.

\begin{figure*}
    \centering
    \includegraphics[width=\textwidth]{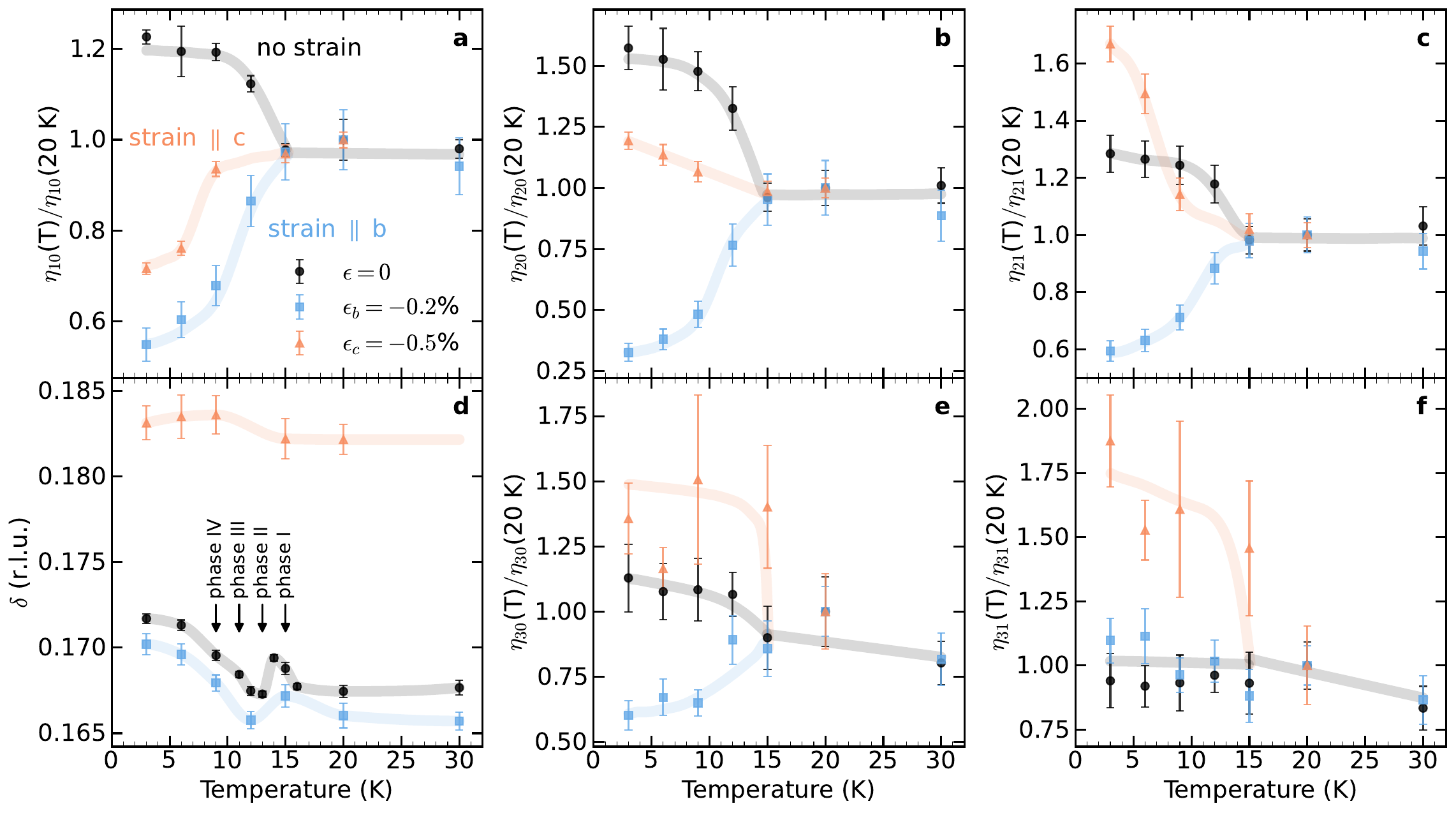}
    \caption{\textbf{Charge localization inside the magnetic phases of \EuAl.} 
    (a-c,e,f) Temperature dependence of $\eta_{10}$, $\eta_{20}$, $\eta_{30}$, $\eta_{21}$ and $\eta_{31}$ for strains (see method section) 
    as indicated, normalized to their respective values at 20~K in the normal state. An increase (decrease) in $\eta_{21}$ and $\eta_{31}$ corresponds to stronger (weaker) charge localization.
    (d) Charge incommensurability, in reciprocal lattice units (r.l.u.), as a function of temperature for the applied strain  as indicated in (a). The arrows show the temperatures where the system enters specific magnetic phases (see main text for description). Solid lines are guides to the eye.
    Error bars denote the standard error of the mean (SEM) extracted from averaging across Brillouin zones.} 
    \label{figure3}
\end{figure*}

 Here we identify EuAl$_4$ as an ideal model system. Prior work has established that this compound hosts charge order that deviates from the simple density wave picture already at elevated temperatures~\cite{kotla_broken_2025a}.  At low temperatures it further develops a cascade  of antiferromagnetic phases with distinct magnetic superstructures driven by the $S = 7/2$ Eu$^{2+}$ moments~\cite{meier_thermodynamic_2022,vibhakar_spontaneous_2024, arai_origin_2026,takagi2022Square,neubauer2025Correlation,nakamura2015Transport,zhu2022Spin}. By systematically studying first, second, and third harmonics of the charge order as a function of temperature, applied magnetic field, and uniaxial pressure (see Fig.~1), we demonstrate a direct coupling between the charge order and the spin order parameters. In the absence of strain and magnetic field, charge localization is enhanced upon entering the magnetically ordered phases. This enhancement in localization is further amplified by applying uniaxial pressure along the $c$-axis.
On the other hand, the in-plane uniaxial pressure exhibits the opposite effect and weakens the charge localization. 
Together, these results establish EuAl$_4$ as a platform for exploring the crossover between density-wave and strongly localized charge order in the presence of long-range magnetic order. The tunable coupling of charge and spin degrees of freedom may also provide a route to controlling the skyrmion phases, with potential implications for spintronics applications.\\

\noindent\textbf{Results}\\
\textit{Charge order in \EuAl:} 
From a \textit{I4/mmm} crystal structure charge ordering  is found at $Q_{co}(n)=\tau \pm n (0,0,\delta)$, where $\tau$ denotes a  fundamental Bragg point, $n$ is an integer, and $\delta$ is the charge incommensurability. In the absence of strain, the charge order onset temperature is $T_{co}=133(1)$~K (see Supplementary Information)  and $\delta(20~\textrm{K})=$ 0.16(7) reciprocal lattice units (r.l.u.) in reasonable agreement with existing literature~\cite{shimomura2019Lattice}.

We find that charge order in \EuAl\ manifests itself through satellite peaks around the Bragg reflections, including the first ($n=1$), second ($n=2$), and even third ($n=3$) harmonics. The first charge order harmonics reach intensities comparable to those of the fundamental Bragg peaks ($n=0$), typically ranging from  one-tenth to one of their intensity across different Brillouin zones. This indicates significant lattice distortions, likely involving the heavy Eu atoms. As shown in Fig.~1, charge order satellites along the reciprocal $c$-axis direction are weaker than those observed in Brillouin zones with a nonzero in-plane momentum component, indicating that the charge ordering involves in-plane lattice distortions.

In what follows, we consider intensity ratios defined by $\eta_{ij}=I_{co}(i)/I_{co}(j)$ (Fig.~1h). With this definition, $\eta_{10}$, $\eta_{20}$, and $\eta_{30}$ compare charge-order satellites of harmonic order $n=1,2,3$  
to the associated Bragg reflection ($n=0$); And $\eta_{21}$ and $\eta_{31}$ describe the ratios between different harmonics.

\textit{Charge localization:} Whereas $\eta_{10}$
is governed by the lattice-distortion amplitude and atomic mass, $\eta_{21}$ and $\eta_{31}$ reflect the degree of charge localization beyond a density-wave order. An ideal charge density wave corresponds to  
$\eta_{21}=\eta_{31}=0$, whereas highly charge-localized dimer systems such as IrTe$_2$~\cite{pascut2014DimerizationInduced,ivashko2017ChargeStripe,rumo2021Insensitivity}, exhibit $\eta_{21}\sim1$ (Fig.~1i). For \EuAl, we generally find $0.01<\eta_{21}<0.1$ and $
0.001<\eta_{31}<0.01$, placing it closer to the density-wave limit. Nevertheless, the finite value of $\eta_{21}$ and $\eta_{31}$ signals a clear deviation from the purely sinusoidal charge-density modulation.

A fundamental question is whether $\eta_{10}$, $\eta_{20}$, and $\eta_{30}$ 
constitute a single order parameter, or instead represent distinct and independent components. In Fig.~1i, it is shown how $\eta_{21}$ and $\eta_{31}$ are strongly temperature dependent --- in contrast to Ir$_{0.95}$Pt$_{0.05}$Te$_2$ where $\eta_{21}$ remains constant. This already indicates that the localization components in EuAl$_4$ act independently. In this study, we report $\eta_{10}$ and $\eta_{21}$ as the system enters its magnetically ordered phases. We bring forward three cases: (1) absence of magnetic field and strain, (2) applied strain in- and out-of-plane, and (3) an applied magnetic field along the $c$-axis. 

Without external stimuli (strain and magnetic field), we plot in Fig.~2 the temperature dependence of $\eta_{10}$, $\eta_{20}$, $\eta_{30}$, $\eta_{21}$, and $\eta_{31}$ normalized to their values at 20~K (above the magnetic ordering temperature).  Upon cooling into the magnetically ordered phases, the charge-order  parameter $\eta_{10}$ increases  by about 20\%, while $\eta_{20}$ exhibits a  
50\% enhancement. This in turn reveals a 30\% enhancement of charge localization measured by $\eta_{21}$. By contrast, $\eta_{31}$ remains essentially constant, independent of temperature.

\textit{Uniaxial pressure application:} Uniaxial pressure applied along an in-plane principal tetragonal axis $\epsilon_b=-0.2$\% reverses the above-mentioned trend. Both $\eta_{10}$ and $\eta_{21}$ are  partially suppressed inside the magnetically ordered phases (Fig.~2), while again $\eta_{31}$ is temperature independent. This observation is likely linked to pressure effects on the magnetically ordered phases as discussed later.

Application of mild (\green{-}0.5\%) compressive $c$-axis strain  
produces striking results. First, in contrast to $b$-axis strain that has  
little effect on the ordering vector, $c$-axis strain enhances the charge incommensurability by about 6\% (Fig.~2d).  To put this into perspective, similar strain application produced a maximum  2\% shift of the stripe order incommensurability in a cuprate compound~\cite{choi_unveiling_2022}. 
Our observation of a large shift in charge-order incommensurability under modest uniaxial pressure  further underscores the fact that even marginal strains can generate  giant electronic responses~\cite{lin_uniaxial_2024}. A second remarkable $c$-axis strain effect is observed upon  entering the magnetically ordered phases: the charge order parameter $\eta_{10}$ is partially suppressed, while the charge localization ratios $\eta_{21}$ and $\eta_{31}$ are enhanced by as much as 70\% at the lowest measured temperature (see Fig.~2).

\textit{Magnetic field dependence:} 
Finally, we discuss the magnetic field dependence in the absence of strain (see Fig.~3). For these experiments a small crystal was used; consequently  only $\eta_{10}$, $\eta_{20}$ and $\eta_{21}$ could be extracted. In the paramagnetic phase (PM), $\eta_{10}$, $\eta_{20}$ and $\eta_{21}$ all display a weak, linear magnetic field dependence. 
The charge order parameter $\eta_{10}$ and the charge localization $\eta_{21}$ increase slightly upon  application of magnetic field. By contrast, within  the magnetically ordered phases, $\eta_{10}$ and $\eta_{21}$ display a much stronger and non-linear magnetic field effect before saturating in the ferromagnetic state (Fig.~3). The charge order incommensurability also displays a magnetic field effect. In the paramagnetic state, the incommensurability is weakly reduced upon application of magnetic field, whereas a  much stronger decrease  of incommensurability is observed across the low-temperature magnetic phases.

To summarize our results, we plot in Fig.~3e and 3f the charge order parameter $\eta_{10}$ and the incommensurability $\delta$ across the low-temperature magnetic phase diagram, where eight different magnetic phases have been identified~\cite{gen_rhombic_2023}. The false color scale is in both cases referenced to the zero-magnetic field paramagnetic state at 20 K. Generally, the low-field magnetic phases partially suppress charge order, indicating a competing interaction. By contrast, the  high-field phases, including the paramagnetic phase,  display an enhancement of charge ordering.

A similar evolution is observed for the incommensurability $\delta$, although with the opposite sign.
At low fields, phases I and IV are characterized by an enlarged incommensurability which is not observed in phases II and III. 
At low temperature, the incommensurability decreases significantly with increasing magnetic field across phases  IV-VIII, before becoming essentially field independent in the high-field normal state (Fig.~3f). Moreover, in the high-temperature normal state, the field effect is much weaker. The strong sensitivity of the ordering vector to the magnetic phase suggests a direct coupling between spin and charge order.\\

\begin{figure*}
    \centering
    \includegraphics[width=\linewidth]{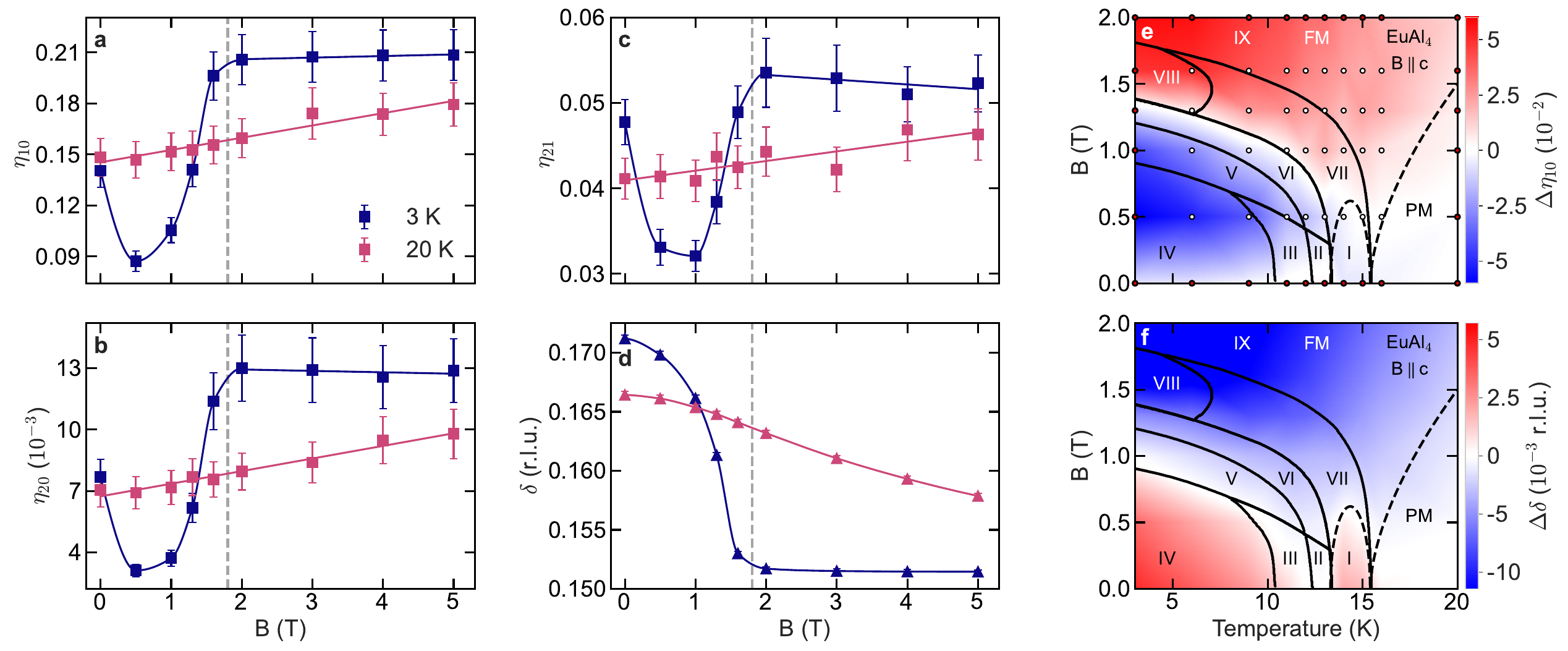}
    \caption{\textbf{Coupling of spin and charge order under magnetic field.} 
    (a-c) Magnetic field  ($B \parallel c$) dependence of the charge localization parameter $\eta_{10},\eta_{20}$ and $\eta_{21}$ at representative temperatures of 3 K and 20 K. (d) Field evolution of the charge order incommensurability $\delta$ for temperatures as indicated. Error bars in (a-d) represent the standard error of the mean. The vertical dashed  line marks the onset of the ferromagnetic ground state. (e,f) display the charge ordering $\eta_{10}$ and the charge incommensurability $\delta$ normalized at 20~K and 0~T, across the temperature-field parameter space, respectively. The false-color profiles are obtained from linear interpolation of a discrete data grid indicated by circles in (e). Solid black lines denote the established phase boundaries \cite{gen_rhombic_2023} separating the various modulated phases. 
   Dashed lines are phase boundaries suggested from this work. In addition to the separation of ferromagnetic (FM), and paramagnetic (PM) states, we propose a separation of the phases --- here labeled --- I and VII.}
   \label{figure2}
\end{figure*}

\noindent\textbf{Discussion}\\
Both uniaxial $c$-axis pressure and chemical substitution tune the tetragonality parameter $a/c$.  Substituting Ba into EuAl$_4$ or SrAl$_4$ reduces the ratio of in-plane to out-of-plane lattice parameters, an effect that is most pronounced in Ba$_{x}$Eu$_{1-x}$Al$_4$~\cite{saraf2025Tuning}. It has also been established that the charge-order incommensurability is inversely correlated with both the onset temperature and the tetragonality ratio~\cite{saraf2025Tuning}.
These systematics  are consistent with our observations under compressive $c$-axis strain, which naturally reduce tetragonality: we find a reduced onset temperature (see Supplementary Information) alongside an enhanced charge incommensurability.
The key advantage of the $c$-axis strain experiment is that it is possible to tune the charge order while keeping the Eu-content --- responsible for the magnetic phases --- unchanged. This makes it possible to study the coupling between complex charge and spin ordering while tuning the charge order.

The magnetic phase diagram of \EuAl\ (Fig.~3e,f) contains at least eight different phases~\cite{meier_thermodynamic_2022}. Among these are spiral and skyrmion phases that are presumably  responsible for intriguing  transport properties such as the anomalous Hall effect~\cite{shang2021Anomalous}. The magnetic phases are characterized by in-plane ordering vectors and are generally understood to be driven by the Eu $S=7/2$ spins. The intensity modulation of the charge-order reflections across Brillouin zones leads us to conclude that the charge ordering also involves the Eu site, and  the diffraction signal further demonstrates that the charge ordering involves in-plane atomic distortions. The coupling between magnetism and charge order is therefore likely rooted in the charge and spin degrees of freedom residing on the same Eu site. A first conclusion is therefore that charge ordering in \EuAl\ goes beyond conventional electron-phonon coupling mechanisms~\cite{korshunov2024Phonon,sukhanov_electronphonon_2025,wang2024Origin}.

In-plane $b$-axis uniaxial strain is known to stabilize a spiral magnetic phase with ordering vector $(\delta_m,0,0)$~\cite{gen_uniaxial_2025}, and increasing strain enhances both the magnetic order parameter and its incommensurability. Our x-ray diffraction experiments show that this evolution is accompanied by a weakening of the charge order parameter and a reduction in charge localization (Fig.~2), pointing to competing ground-state energetics between the spiral magnetic phase and the charge-ordered state. By contrast, compressive $c$-axis strain enhances charge localization within the magnetically ordered phase, strongly suggesting that certain magnetic structures are compatible with — and possibly cooperative with — charge order. In such phases, charge and spin degrees of freedom  act collaboratively rather than competitively. Determining whether this cooperative magnetic phase is chiral will require Hall effect~\cite{shang2021Anomalous}, neutron scattering~\cite{gen_uniaxial_2025}, or polarised light scattering~\cite{vibhakar_spontaneous_2024} measurements under $c$-axis strain. More broadly, our results establish a framework in which the relationship between charge and spin order in \EuAl\ is not invariant, but depends sensitively on the symmetry of the applied perturbation --- a finding with direct implications for the control of topological magnetic textures in this family of materials.\\

\noindent\textbf{Methods}\\
Single crystals of EuAl$_4$ were grown using a self-flux method. High-purity Eu (4N, vacuum-distilled) and Al (4N) were loaded into pre-baked alumina crucibles in a molar ratio of Eu:Al = 1:10 and sealed in a quartz tube. The mixture was heated to 1030~$^\circ$C, with occasional shaking to ensure homogenization. It was then cooled to 700~$^\circ$C at a rate of 2~$^\circ$C/h, after which the remaining flux was removed by centrifugation, yielding shiny, well-faceted single crystals. Phase purity and crystal structure were confirmed by powder x-ray  diffraction, shown in the supplementary.

High-energy x-ray diffraction experiments were carried out at the P21.1 beamline~\cite{v.zimmermann_p211_2025} at the PETRA-III synchrotron using 102~keV photons.
A horizontal 10-T cryomagnet~\cite{hucker_enhanced_2013,christensen_bulk_2014} was used to apply magnetic field along the $c$-axis. A uniaxial pressure cell~\cite{choi_unveiling_2022,simutis_singledomain_2022} allowing strain application perpendicular to the horizontal scattering plane was employed. Single crystals of EuAl$_4$ were aligned so that pressure is applied either along the crystallographic $b$- or $c$-axis. Compressive strain $\epsilon_b=(b-b_0)/b_0$ and $\epsilon_c=(c-c_0)/c_0$ along the $b$- and $c$-axis are derived from the lattice parameters $b$ and $c$ obtained from the fundamental Bragg points $\tau$ (see Supplementary Information). The zero-strain lattice parameters $a_0=b_0=4.38$~\AA\ and $c_0=11.19$~\AA\ are consistent with previous  reports~\cite{korshunov2024Phonon}. In this study (due to limited beam time allocation), we use magnetic field and uniaxial pressure as separate external tuning parameters, applying them individually rather than in combination.
Diffracted intensity was recorded using a Pilatus3 X CdTe 2M x-ray detector.

Crystal growth and x-ray diffraction methodology for the Ir$_{0.95}$Pt$_{0.05}$Te$_2$ results are described in Ref.~\cite{ivashko2017ChargeStripe}.\\

\noindent\textbf{Acknowledgments}\\
We thank Jiatu Liu for assistance with data processing and Leonardo Martinelli and Izabela Biało for advices during sample preparation. The work at CUHK is supported by the Research Grants Council of Hong Kong (CUHK 24306223), the Guangdong Provincial Quantum Science Strategic Initiative (GDZX2401012), CUHK Direct Grant (4053671, 4053791), and the 1+1+1 CUHK–CUHK(SZ)–GDSTC Joint Collaboration Fund (2025A0505000079). X.H. acknowledges support from UZH Candoc Grant (FK-25-088).
M. H. acknowledges funding support from the Helmsley Charitable Trust grant \#2112-04911.  Work at the University of Zagreb was supported by Croatian Science Foundation projects IP-2025-02-3955, IP-2025-02-1189, and IP-2022-10-3382; the CeNIKS project, co-financed by the Croatian Government and the European Union through the European Regional Development Fund under the Competitiveness and Cohesion Operational Programme (Grant No. KK.01.1.1.02.0013); and the NextGenerationEU project, as part of the National Recovery and Resilience Plan 2021–2026, through institutional grants of the University of Zagreb Faculty of Science (ProPuBFO, NQMat).
The work at TU Wien was supported by the Austrian Science Fund (FWF) [10.55776/F86; 10.55776/P35945].
We acknowledge DESY (Hamburg, Germany), a member of the Helmholtz Association HGF, for the provision of experimental facilities.
Parts of this research were carried out at beamline P21.1 at PETRA III. Beamtime was allocated for proposal 20251089.\\

\noindent\textbf{Author contributions}\\
M.B., X.H. and T.W. contributed equally. Q.W. and J.C. conceived the project. T.S., P.S., A.A., N.B. and M.N., grew the \EuAl\ crystals and S.P., K.K., and M.N. grew the Ir$_{1-x}$Pt$_x$Te$_2$ crystals.
T.W., F.Y., Y.W. and J.M. prepared the XRD magnetic field experiments. M.B., J.K., and X.H. prepared the uniaxial pressure experiments. The experiments on \EuAl\ were carried out by T.W., M.B., X.H., F.I. and M.v.Z. and the experiments on Ir$_{1-x}$Pt$_x$Te$_2$ were carried out by O.I., M.H., J.C., and M.v.Z. Data analysis was carried out by T.W., M.B., J.O. and X.H. Manuscript was written by Q.W. and J.C. with input from all authors. \\
	
\noindent\textbf{Data availability} \\
Data supporting the findings of this study are available from the corresponding authors upon request. Source data are deposited in a Zenodo repository: \hyperlink{https://doi.org/10.5281/zenodo.20426014}{https://doi.org/10.5281/zenodo.20426014}.\\

\noindent\textbf{Competing interests} \\
The authors declare no competing interests.\\

\bibliography{Johan-LQMR, extra_bib}

@article{kiss2007Chargeordermaximized,
	title = {Charge-order-maximized momentum-dependent superconductivity},
	volume = {3},
	copyright = {http://www.springer.com/tdm},
	issn = {1745-2473, 1745-2481},
	url = {https://www.nature.com/articles/nphys699},
	doi = {10.1038/nphys699},
	language = {en},
	number = {10},
	urldate = {2026-05-01},
	journal = {Nature Physics},
	author = {Kiss, T. and Yokoya, T. and Chainani, A. and Shin, S. and Hanaguri, T. and Nohara, M. and Takagi, H.},
	month = oct,
	year = {2007},
	pages = {720--725},
}

@article{pascut2014DimerizationInduced,
	title = {Dimerization-{Induced} {Cross}-{Layer} {Quasi}-{Two}-{Dimensionality} in {Metallic} {IrTe} 2},
	volume = {112},
	copyright = {http://link.aps.org/licenses/aps-default-license},
	issn = {0031-9007, 1079-7114},
	url = {https://link.aps.org/doi/10.1103/PhysRevLett.112.086402},
	doi = {10.1103/PhysRevLett.112.086402},
	language = {en},
	number = {8},
	urldate = {2026-05-01},
	journal = {Physical Review Letters},
	author = {Pascut, G. L. and Haule, K. and Gutmann, M. J. and Barnett, S. A. and Bombardi, A. and Artyukhin, S. and Birol, T. and Vanderbilt, D. and Yang, J. J. and Cheong, S.-W. and Kiryukhin, V.},
	month = feb,
	year = {2014},
	pages = {086402},
}

@article{ivashko2017ChargeStripe,
	title = {Charge-{Stripe} {Order} and {Superconductivity} in {Ir1}−{xPtxTe2}},
	volume = {7},
	issn = {2045-2322},
	url = {https://www.nature.com/articles/s41598-017-16945-7},
	doi = {10.1038/s41598-017-16945-7},
	language = {en},
	number = {1},
	urldate = {2026-05-01},
	journal = {Scientific Reports},
	author = {Ivashko, O. and Yang, L. and Destraz, D. and Martino, E. and Chen, Y. and Guo, C. Y. and Yuan, H. Q. and Pisoni, A. and Matus, P. and Pyon, S. and Kudo, K. and Nohara, M. and Forró, L. and Rønnow, H. M. and Hücker, M. and V. Zimmermann, M. and Chang, J.},
	month = dec,
	year = {2017},
	pages = {17157},
}

@article{revelli2019Resonant,
	title = {Resonant inelastic x-ray incarnation of {Young}’s double-slit experiment},
	volume = {5},
	issn = {2375-2548},
	url = {https://www.science.org/doi/10.1126/sciadv.aav4020},
	doi = {10.1126/sciadv.aav4020},
	language = {en},
	number = {1},
	urldate = {2026-05-01},
	journal = {Science Advances},
	author = {Revelli, A. and Moretti Sala, M. and Monaco, G. and Becker, P. and Bohatý, L. and Hermanns, M. and Koethe, T. C. and Fröhlich, T. and Warzanowski, P. and Lorenz, T. and Streltsov, S. V. and Van Loosdrecht, P. H. M. and Khomskii, D. I. and Van Den Brink, J. and Grüninger, M.},
	month = jan,
	year = {2019},
	pages = {eaav4020},
}

@article{fawcett1988Spindensitywave,
	title = {Spin-density-wave antiferromagnetism in chromium},
	volume = {60},
	copyright = {http://link.aps.org/licenses/aps-default-license},
	issn = {0034-6861},
	url = {https://link.aps.org/doi/10.1103/RevModPhys.60.209},
	doi = {10.1103/RevModPhys.60.209},
	language = {en},
	number = {1},
	urldate = {2026-05-01},
	journal = {Reviews of Modern Physics},
	author = {Fawcett, Eric},
	month = jan,
	year = {1988},
	pages = {209--283},
}

@article{abbamonte2005Spatially,
	title = {Spatially modulated '{Mottness}' in {La2}-{xBaxCuO4}},
	volume = {1},
	issn = {1745-2473, 1745-2481},
	url = {https://www.nature.com/articles/nphys178},
	doi = {10.1038/nphys178},
	language = {en},
	number = {3},
	urldate = {2026-05-01},
	journal = {Nature Physics},
	author = {Abbamonte, P. and Rusydi, A. and Smadici, S. and Gu, G. D. and Sawatzky, G. A. and Feng, D. L.},
	month = dec,
	year = {2005},
	pages = {155--158},
}

@article{shang2021Anomalous,
	title = {Anomalous {Hall} resistivity and possible topological {Hall} effect in the {EuAl} 4 antiferromagnet},
	volume = {103},
	issn = {2469-9950, 2469-9969},
	url = {https://link.aps.org/doi/10.1103/PhysRevB.103.L020405},
	doi = {10.1103/PhysRevB.103.L020405},
	language = {en},
	number = {2},
	urldate = {2026-05-01},
	journal = {Physical Review B},
	author = {Shang, T. and Xu, Y. and Gawryluk, D. J. and Ma, J. Z. and Shiroka, T. and Shi, M. and Pomjakushina, E.},
	month = jan,
	year = {2021},
	pages = {L020405},
}

@article{shimomura2019Lattice,
	title = {Lattice {Modulation} and {Structural} {Phase} {Transition} in the {Antiferromagnet} {EuAl}$_{\textrm{4}}$},
	volume = {88},
	issn = {0031-9015, 1347-4073},
	url = {https://journals.jps.jp/doi/10.7566/JPSJ.88.014602},
	doi = {10.7566/JPSJ.88.014602},
	language = {en},
	number = {1},
	urldate = {2026-05-02},
	journal = {Journal of the Physical Society of Japan},
	author = {Shimomura, Susumu and Murao, Hiroki and Tsutsui, Satoshi and Nakao, Hironori and Nakamura, Ai and Hedo, Masato and Nakama, Takao and Ōnuki, Yoshichika},
	month = jan,
	year = {2019},
	pages = {014602},
}

@article{rumo2021Insensitivity,
	title = {Insensitivity of the striped charge orders in {IrTe} 2 to alkali surface doping implies their structural origin},
	volume = {5},
	issn = {2475-9953},
	url = {https://link.aps.org/doi/10.1103/PhysRevMaterials.5.074002},
	doi = {10.1103/PhysRevMaterials.5.074002},
	language = {en},
	number = {7},
	urldate = {2026-05-02},
	journal = {Physical Review Materials},
	author = {Rumo, M. and Pulkkinen, A. and Salzmann, B. and Kremer, G. and Hildebrand, B. and Ma, K. Y. and Von Rohr, F. O. and Nicholson, C. W. and Jaouen, T. and Monney, C.},
	month = jul,
	year = {2021},
	pages = {074002},
}

@article{xi2015Strongly,
	title = {Strongly enhanced charge-density-wave order in monolayer {NbSe2}},
	volume = {10},
	issn = {1748-3387, 1748-3395},
	url = {https://www.nature.com/articles/nnano.2015.143},
	doi = {10.1038/nnano.2015.143},
	language = {en},
	number = {9},
	urldate = {2026-05-02},
	journal = {Nature Nanotechnology},
	author = {Xi, Xiaoxiang and Zhao, Liang and Wang, Zefang and Berger, Helmuth and Forró, László and Shan, Jie and Mak, Kin Fai},
	month = sep,
	year = {2015},
	pages = {765--769},
}

@article{kundu2024Charge,
	title = {Charge density waves and the effects of uniaxial strain on the electronic structure of {2H}-{NbSe2}},
	volume = {5},
	issn = {2662-4443},
	url = {https://www.nature.com/articles/s43246-024-00661-7},
	doi = {10.1038/s43246-024-00661-7},
	language = {en},
	number = {1},
	urldate = {2026-05-02},
	journal = {Communications Materials},
	author = {Kundu, Asish K. and Rajapitamahuni, Anil and Vescovo, Elio and Klimovskikh, Ilya I. and Berger, Helmuth and Valla, Tonica},
	month = oct,
	year = {2024},
	pages = {208},
}

@article{teng2022Discovery,
	title = {Discovery of charge density wave in a kagome lattice antiferromagnet},
	volume = {609},
	issn = {0028-0836, 1476-4687},
	url = {https://www.nature.com/articles/s41586-022-05034-z},
	doi = {10.1038/s41586-022-05034-z},
	language = {en},
	number = {7927},
	urldate = {2026-05-03},
	journal = {Nature},
	author = {Teng, Xiaokun and Chen, Lebing and Ye, Feng and Rosenberg, Elliott and Liu, Zhaoyu and Yin, Jia-Xin and Jiang, Yu-Xiao and Oh, Ji Seop and Hasan, M. Zahid and Neubauer, Kelly J. and Gao, Bin and Xie, Yaofeng and Hashimoto, Makoto and Lu, Donghui and Jozwiak, Chris and Bostwick, Aaron and Rotenberg, Eli and Birgeneau, Robert J. and Chu, Jiun-Haw and Yi, Ming and Dai, Pengcheng},
	month = sep,
	year = {2022},
	pages = {490--495},
}

@article{takagi2022Square,
	title = {Square and rhombic lattices of magnetic skyrmions in a centrosymmetric binary compound},
	volume = {13},
	issn = {2041-1723},
	url = {https://www.nature.com/articles/s41467-022-29131-9},
	doi = {10.1038/s41467-022-29131-9},
	language = {en},
	number = {1},
	urldate = {2026-05-03},
	journal = {Nature Communications},
	author = {Takagi, Rina and Matsuyama, Naofumi and Ukleev, Victor and Yu, Le and White, Jonathan S. and Francoual, Sonia and Mardegan, José R. L. and Hayami, Satoru and Saito, Hiraku and Kaneko, Koji and Ohishi, Kazuki and Ōnuki, Yoshichika and Arima, Taka-hisa and Tokura, Yoshinori and Nakajima, Taro and Seki, Shinichiro},
	month = mar,
	year = {2022},
	pages = {1472},
}

@article{neubauer2025Correlation,
	title = {Correlation between complex spin textures and the magnetocaloric and {Hall} effects in {Eu} ( {Ga} 1 − x {Al} x ) 4 ( x = 0.9 , 1)},
	volume = {111},
	issn = {2469-9950, 2469-9969},
	url = {https://link.aps.org/doi/10.1103/PhysRevB.111.165136},
	doi = {10.1103/PhysRevB.111.165136},
	language = {en},
	number = {16},
	urldate = {2026-05-03},
	journal = {Physical Review B},
	author = {Neubauer, Kelly J. and Allen, Kevin and Moya, Jaime M. and Klemm, Mason L. and Ye, Feng and Morgan, Zachary and DeBeer-Schmitt, Lisa and Tian, Wei and Morosan, Emilia and Dai, Pengcheng},
	month = apr,
	year = {2025},
	pages = {165136},
}

@article{zhu2022Spin,
	title = {Spin order and fluctuations in the {EuAl} 4 and {EuGa} 4 topological antiferromagnets: {A} μ {SR} study},
	volume = {105},
	issn = {2469-9950, 2469-9969},
	shorttitle = {Spin order and fluctuations in the {EuAl} 4 and {EuGa} 4 topological antiferromagnets},
	url = {https://link.aps.org/doi/10.1103/PhysRevB.105.014423},
	doi = {10.1103/PhysRevB.105.014423},
	language = {en},
	number = {1},
	urldate = {2026-05-03},
	journal = {Physical Review B},
	author = {Zhu, X. Y. and Zhang, H. and Gawryluk, D. J. and Zhen, Z. X. and Yu, B. C. and Ju, S. L. and Xie, W. and Jiang, D. M. and Cheng, W. J. and Xu, Y. and Shi, M. and Pomjakushina, E. and Zhan, Q. F. and Shiroka, T. and Shang, T.},
	month = jan,
	year = {2022},
	pages = {014423},
}

@article{nakamura2015Transport,
	title = {Transport and {Magnetic} {Properties} of {EuAl}$_{\textrm{4}}$ and {EuGa}$_{\textrm{4}}$},
	volume = {84},
	issn = {0031-9015, 1347-4073},
	url = {http://journals.jps.jp/doi/10.7566/JPSJ.84.124711},
	doi = {10.7566/JPSJ.84.124711},
	language = {en},
	number = {12},
	urldate = {2026-05-03},
	journal = {Journal of the Physical Society of Japan},
	author = {Nakamura, Ai and Uejo, Taro and Honda, Fuminori and Takeuchi, Tetsuya and Harima, Hisatomo and Yamamoto, Etsuji and Haga, Yoshinori and Matsubayashi, Kazuyuki and Uwatoko, Yoshiya and Hedo, Masato and Nakama, Takao and Ōnuki, Yoshichika},
	month = dec,
	year = {2015},
	pages = {124711},
}

@article{korshunov2024Phonon,
	title = {Phonon softening and atomic modulations in {EuAl} 4},
	volume = {110},
	issn = {2469-9950, 2469-9969},
	url = {https://link.aps.org/doi/10.1103/PhysRevB.110.045102},
	doi = {10.1103/PhysRevB.110.045102},
	language = {en},
	number = {4},
	urldate = {2026-05-03},
	journal = {Physical Review B},
	author = {Korshunov, A. N. and Sukhanov, A. S. and Gebel, S. and Pavlovskii, M. S. and Andriushin, N. D. and Gao, Y. and Moya, J. M. and Morosan, E. and Rahn, M. C.},
	month = jul,
	year = {2024},
	pages = {045102},
}

@article{wang2024Origin,
	title = {Origin of charge density wave in topological semimetals {SrAl4} and {EuAl4}},
	volume = {7},
	issn = {2399-3650},
	url = {https://www.nature.com/articles/s42005-024-01600-1},
	doi = {10.1038/s42005-024-01600-1},
	language = {en},
	number = {1},
	urldate = {2026-05-03},
	journal = {Communications Physics},
	author = {Wang, Lin-Lin and Nepal, Niraj K. and Canfield, Paul C.},
	month = mar,
	year = {2024},
	pages = {111},
}

@article{radaelli2002Formation,
	title = {Formation of isomorphic {Ir3}+ and {Ir4}+ octamers and spin dimerization in the spinel {CuIr2S4}},
	volume = {416},
	copyright = {http://www.springer.com/tdm},
	issn = {0028-0836, 1476-4687},
	url = {https://www.nature.com/articles/416155a},
	doi = {10.1038/416155a},
	language = {en},
	number = {6877},
	urldate = {2026-05-03},
	journal = {Nature},
	author = {Radaelli, Paolo G. and Horibe, Y. and Gutmann, Matthias J. and Ishibashi, Hiroki and Chen, C. H. and Ibberson, Richard M. and Koyama, Y. and Hor, Yew-San and Kiryukhin, Valery and Cheong, Sang-Wook},
	month = mar,
	year = {2002},
	pages = {155--158},
}

@article{hwang2022Largegap,
	title = {Large-gap insulating dimer ground state in monolayer {IrTe2}},
	volume = {13},
	issn = {2041-1723},
	url = {https://www.nature.com/articles/s41467-022-28542-y},
	doi = {10.1038/s41467-022-28542-y},
	language = {en},
	number = {1},
	urldate = {2026-05-03},
	journal = {Nature Communications},
	author = {Hwang, Jinwoong and Kim, Kyoo and Zhang, Canxun and Zhu, Tiancong and Herbig, Charlotte and Kim, Sooran and Kim, Bongjae and Zhong, Yong and Salah, Mohamed and El-Desoky, Mohamed M. and Hwang, Choongyu and Shen, Zhi-Xun and Crommie, Michael F. and Mo, Sung-Kwan},
	month = feb,
	year = {2022},
	pages = {906},
}

@article{katukuri2022Charge,
	title = {Charge ordering in {Ir} dimers in the ground state of {Ba} 5 {AlIr} 2 {O} 11},
	volume = {105},
	issn = {2469-9950, 2469-9969},
	url = {https://link.aps.org/doi/10.1103/PhysRevB.105.075114},
	doi = {10.1103/PhysRevB.105.075114},
	language = {en},
	number = {7},
	urldate = {2026-05-03},
	journal = {Physical Review B},
	author = {Katukuri, Vamshi M. and Lu, Xingye and McNally, D. E. and Dantz, Marcus and Strocov, Vladimir N. and Sala, M. Moretti and Upton, M. H. and Terzic, J. and Cao, G. and Yazyev, Oleg V. and Schmitt, Thorsten},
	month = feb,
	year = {2022},
	pages = {075114},
}

@article{magnaterra2023RIXS,
	title = {{RIXS} interferometry and the role of disorder in the quantum magnet {Ba} 3 {Ti} 3 − x {Ir} x {O} 9},
	volume = {5},
	issn = {2643-1564},
	url = {https://link.aps.org/doi/10.1103/PhysRevResearch.5.013167},
	doi = {10.1103/PhysRevResearch.5.013167},
	language = {en},
	number = {1},
	urldate = {2026-05-03},
	journal = {Physical Review Research},
	author = {Magnaterra, M. and Moretti Sala, M. and Monaco, G. and Becker, P. and Hermanns, M. and Warzanowski, P. and Lorenz, T. and Khomskii, D. I. and Van Loosdrecht, P. H. M. and Van Den Brink, J. and Grüninger, M.},
	month = mar,
	year = {2023},
	pages = {013167},
}

@article{kotla_broken_2025a,
	title = {Broken inversion symmetry in the charge density wave phase in {EuAl} 4},
	volume = {112},
	issn = {2469-9950, 2469-9969},
	url = {https://link.aps.org/doi/10.1103/kl2z-brms},
	doi = {10.1103/kl2z-brms},
	language = {en},
	number = {6},
	urldate = {2026-04-30},
	journal = {Physical Review B},
	author = {Kotla, Surya Rohith and Noohinejad, Leila and Pokhriyal, Preeti and Tolkiehn, Martin and Agarwal, Harshit and Ramakrishnan, Sitaram and Van Smaalen, Sander},
	month = aug,
	year = {2025},
	pages = {064113},
}

@article{saraf2025Tuning,
	title = {Tuning incommensurate charge order in {Ba} 1 − x {Sr} x {Al} 4 and {Ba} 1 − y {Eu} y {Al} 4},
	volume = {112},
	issn = {2469-9950, 2469-9969},
	url = {https://link.aps.org/doi/10.1103/d8yy-1w3l},
	doi = {10.1103/d8yy-1w3l},
	language = {en},
	number = {3},
	urldate = {2026-04-20},
	journal = {Physical Review B},
	author = {Saraf, Prathum and Clements, Eleanor M. and Sokratov, Danila and Saha, Shanta and Zavalij, Peter and Heitmann, Thomas W. and Lynn, Jeffrey W. and Bernal-Choban, Camille and Chaudhuri, Dipanjan and Kengle, Caitlin S. and Su, Yue and Bettler, Simon and Manning, Nathan and Abbamonte, Peter and Biswas, Sananda and Valentí, Roser and Paglione, Johnpierre},
	month = jul,
	year = {2025},
	pages = {035151},
}

@article{arai_origin_2026,
	title = {Origin of multiple skyrmion phases in {EuAl4}},
	volume = {17},
	issn = {2041-1723},
	url = {https://www.nature.com/articles/s41467-026-71020-y},
	doi = {10.1038/s41467-026-71020-y},
	language = {en},
	number = {1},
	urldate = {2026-04-15},
	journal = {Nature Communications},
	author = {Arai, Yuki and Nakayama, Kosuke and Honma, Asuka and Souma, Seigo and Shiga, Daisuke and Kumigashira, Hiroshi and Takahashi, Takashi and Segawa, Kouji and Sato, Takafumi},
	month = apr,
	year = {2026},
	pages = {3162},
}

@article{vibhakar_spontaneous_2024,
	title = {Spontaneous reversal of spin chirality and competing phases in the topological magnet {EuAl}$_{\textrm{4}}$},
	volume = {7},
	issn = {2399-3650},
	url = {https://www.nature.com/articles/s42005-024-01802-7},
	doi = {10.1038/s42005-024-01802-7},
	number = {1},
	urldate = {2025-09-30},
	journal = {Communications Physics},
	author = {Vibhakar, Anuradha M. and Khalyavin, Dmitry D. and Orlandi, Fabio and Moya, Jamie M. and Lei, Shiming and Morosan, Emilia and Bombardi, Alessandro},
	month = sep,
	year = {2024},
	pages = {313},
}

@article{choi_unveiling_2022,
	title = {Unveiling {Unequivocal} {Charge} {Stripe} {Order} in a {Prototypical} {Cuprate} {Superconductor}},
	volume = {128},
	url = {https://link.aps.org/doi/10.1103/PhysRevLett.128.207002},
	doi = {10.1103/PhysRevLett.128.207002},
	number = {20},
	urldate = {2024-03-24},
	journal = {Physical Review Letters},
	publisher = {American Physical Society},
	author = {Choi, J. and Wang, Q. and Jöhr, S. and Christensen, N. B. and Küspert, J. and Bucher, D. and Biscette, D. and Fischer, M. H. and Hücker, M. and Kurosawa, T. and Momono, N. and Oda, M. and Ivashko, O. and Zimmermann, M. v. and Janoschek, M. and Chang, J.},
	month = may,
	year = {2022},
	pages = {207002},
}

@article{lin_uniaxial_2024,
	title = {Uniaxial strain tuning of charge modulation and singularity in a kagome superconductor},
	volume = {15},
	copyright = {2024 The Author(s)},
	issn = {2041-1723},
	url = {https://www.nature.com/articles/s41467-024-53737-w},
	doi = {10.1038/s41467-024-53737-w},
	language = {en},
	number = {1},
	urldate = {2026-01-04},
	journal = {Nature Communications},
	publisher = {Nature Publishing Group},
	author = {Lin, Chun and Consiglio, Armando and Forslund, Ola Kenji and Küspert, Julia and Denner, M. Michael and Lei, Hechang and Louat, Alex and Watson, Matthew D. and Kim, Timur K. and Cacho, Cephise and Carbone, Dina and Leandersson, Mats and Polley, Craig and Balasubramanian, Thiagarajan and Sante, Domenico Di and Thomale, Ronny and Guguchia, Zurab and Sangiovanni, Giorgio and Neupert, Titus and Chang, Johan},
	month = dec,
	year = {2024},
	pages = {10466},
}

@misc{gen_uniaxial_2025,
	title = {Uniaxial stress control of versatile helimagnetic phases in the square-lattice itinerant magnet {EuAl}\$\_\{4\}\$},
	copyright = {Creative Commons Attribution 4.0 International},
	url = {https://arxiv.org/abs/2511.07079},
	doi = {10.48550/ARXIV.2511.07079},
	urldate = {2025-11-12},
	publisher = {arXiv},
	author = {Gen, Masaki and Nomoto, Takuya and Saito, Hiraku and Nakajima, Taro and Tokunaga, Yusuke and Takagi, Rina and Seki, Shinichiro and Arima, Taka-hisa},
	year = {2025},
	note = {Version Number: 1},
}

@article{meier_thermodynamic_2022,
	title = {Thermodynamic insights into the intricate magnetic phase diagram of {EuAl} 4},
	volume = {106},
	issn = {2469-9950, 2469-9969},
	url = {https://link.aps.org/doi/10.1103/PhysRevB.106.094421},
	doi = {10.1103/PhysRevB.106.094421},
	language = {en},
	number = {9},
	urldate = {2025-09-30},
	journal = {Physical Review B},
	author = {Meier, William R. and Torres, James R. and Hermann, Raphael P. and Zhao, Jiyong and Lavina, Barbara and Sales, Brian C. and May, Andrew F.},
	month = sep,
	year = {2022},
	pages = {094421},
}

@article{gruner_dynamics_1988,
	title = {The dynamics of charge-density waves},
	volume = {60},
	doi = {10.1103/RevModPhys.60.1129},
	number = {4},
	journal = {Rev. Mod. Phys.},
	author = {Grüner, G.},
	month = oct,
	year = {1988},
	pages = {1129--1181},
}

@article{simutis_singledomain_2022,
	title = {Single-domain stripe order in a high-temperature superconductor},
	volume = {5},
	copyright = {2022 The Author(s)},
	issn = {2399-3650},
	url = {https://www.nature.com/articles/s42005-022-01061-4},
	doi = {10.1038/s42005-022-01061-4},
	number = {1},
	urldate = {2024-03-24},
	journal = {Communications Physics},
	author = {Simutis, Gediminas and Küspert, Julia and Wang, Qisi and Choi, Jaewon and Bucher, Damian and Boehm, Martin and Bourdarot, Frédéric and Bertelsen, Mads and Wang, Chennan N. and Kurosawa, Tohru and Momono, Naoki and Oda, Migaku and Månsson, Martin and Sassa, Yasmine and Janoschek, Marc and Christensen, Niels B. and Chang, Johan and Mazzone, Daniel G.},
	month = nov,
	year = {2022},
	pages = {1--7},
}

@article{fink_phase_2011,
	title = {Phase diagram of charge order in {La}\_1.8{\textbackslash}ensuremath-{xEu}\_0.{2Sr}\_xCuO\_4 from resonant soft x-ray diffraction},
	volume = {83},
	url = {https://link.aps.org/doi/10.1103/PhysRevB.83.092503},
	doi = {10.1103/PhysRevB.83.092503},
	number = {9},
	journal = {Phys. Rev. B},
	publisher = {American Physical Society},
	author = {Fink, Jörg and Soltwisch, Victor and Geck, Jochen and Schierle, Enrico and Weschke, Eugen and Büchner, Bernd},
	month = mar,
	year = {2011},
	pages = {092503},
}

@article{v.zimmermann_p211_2025,
	title = {P21.1 at {PETRA} {III} – a high-energy {X}-ray diffraction beamline for physics and chemistry},
	volume = {32},
	copyright = {https://creativecommons.org/licenses/by/4.0/},
	issn = {1600-5775},
	url = {https://journals.iucr.org/s/issues/2025/03/00/yi5172/},
	doi = {10.1107/S1600577525002826},
	language = {en},
	number = {3},
	urldate = {2025-05-29},
	journal = {Journal of Synchrotron Radiation},
	author = {v. Zimmermann, M. and Ivashko, O. and Igoa Saldaña, F. and Liu, J. and Glaevecke, P. and Gutowski, O. and Nowak, R. and Köhler, K. and Winkler, B. and Schöps, A. and Schulte-Schrepping, H. and Dippel, A.-C.},
	month = may,
	year = {2025},
	pages = {802--814},
}

@article{vonarx_fate_2023,
	title = {Fate of charge order in overdoped {La}-based cuprates},
	volume = {8},
	copyright = {2023 The Author(s)},
	issn = {2397-4648},
	url = {https://www.nature.com/articles/s41535-023-00539-w},
	doi = {10.1038/s41535-023-00539-w},
	language = {en},
	number = {1},
	urldate = {2025-08-11},
	journal = {npj Quantum Materials},
	publisher = {Nature Publishing Group},
	author = {von Arx, K. and Wang, Qisi and Mustafi, S. and Mazzone, D. G. and Horio, M. and Mukkattukavil, D. John and Pomjakushina, E. and Pyon, S. and Takayama, T. and Takagi, H. and Kurosawa, T. and Momono, N. and Oda, M. and Brookes, N. B. and Betto, D. and Zhang, W. and Asmara, T. C. and Tseng, Y. and Schmitt, T. and Sassa, Y. and Chang, J.},
	month = jan,
	year = {2023},
	pages = {7},
}

@article{tranquada_evidence_1995,
	title = {Evidence for stripe correlations of spins and holes in copper oxide superconductors},
	volume = {375},
	url = {https://www.nature.com/articles/375561a0},
	journal = {Nature},
	author = {Tranquada, J. M. and Sternlieb, B. J. and Axe, J. D. and Nakamura, Y. and Uchida, S.},
	year = {1995},
	pages = {561},
}

@article{hucker_enhanced_2013,
	title = {Enhanced charge stripe order of superconducting {La}$_{\textrm{2-x}}${Ba}$_{\textrm{x}}${CuO}$_{\textrm{4}}$ in a magnetic field},
	volume = {87},
	url = {https://link.aps.org/doi/10.1103/PhysRevB.87.014501},
	doi = {10.1103/PhysRevB.87.014501},
	number = {1},
	journal = {Phys. Rev. B},
	author = {Hücker, M. and v. Zimmermann, M. and Xu, Z. J. and Wen, J. S. and Gu, G. D. and Tranquada, J. M.},
	month = jan,
	year = {2013},
	pages = {014501},
}

@article{sukhanov_electronphonon_2025,
	title = {Electron-phonon coupling in {EuAl} 4 under hydrostatic pressure},
	volume = {111},
	issn = {2469-9950, 2469-9969},
	url = {https://link.aps.org/doi/10.1103/PhysRevB.111.195150},
	doi = {10.1103/PhysRevB.111.195150},
	language = {en},
	number = {19},
	urldate = {2026-02-17},
	journal = {Physical Review B},
	author = {Sukhanov, A. S. and Gebel, S. and Korshunov, A. N. and Andriushin, N. D. and Pavlovskii, M. S. and Gao, Y. and Moya, J. M. and Allen, K. and Morosan, E. and Rahn, M. C.},
	month = may,
	year = {2025},
	pages = {195150},
}

@article{monceau_electronic_2012,
	title = {Electronic crystals: an experimental overview},
	volume = {61},
	journal = {Advances in Physics},
	author = {Monceau, Pierre},
	year = {2012},
	pages = {325--581},
}

@article{senn_charge_2012,
	title = {Charge order and three-site distortions in the {Verwey} structure of magnetite},
	volume = {481},
	copyright = {http://www.springer.com/tdm},
	issn = {0028-0836, 1476-4687},
	url = {https://www.nature.com/articles/nature10704},
	doi = {10.1038/nature10704},
	language = {en},
	number = {7380},
	urldate = {2025-07-14},
	journal = {Nature},
	publisher = {Springer Science and Business Media LLC},
	author = {Senn, Mark S. and Wright, Jon P. and Attfield, J. Paul},
	month = jan,
	year = {2012},
	pages = {173--176},
}

@article{christensen_bulk_2014,
	title = {Bulk charge stripe order competing with superconductivity in {La}$_{\textrm{2}}${Sr}$_{\textrm{x}}${CuO}$_{\textrm{4}}$ (x=0.12)},
	url = {https://arxiv.org/abs/1404.3192},
	doi = {10.48550/arXiv.1404.3192},
	journal = {arXiv:1404.3192},
	author = {Christensen, N. B. and Chang, J. and Larsen, J. and Fujita, M. and Oda, M. and Ido, M. and Momono, N. and Forgan, E. M. and Holmes, A. T. and Mesot, J. and Huecker, M. and Zimmermann, M. v},
	year = {2014},
}

@article{jaramillo_breakdown_2009a,
	title = {Breakdown of the {Bardeen}–{Cooper}–{Schrieffer} ground state at a quantum phase transition},
	volume = {459},
	issn = {1476-4687},
	url = {http://dx.doi.org/10.1038/nature08008},
	doi = {10.1038/nature08008},
	number = {7245},
	journal = {Nature},
	publisher = {Springer Science and Business Media LLC},
	author = {Jaramillo, R. and Feng, Yejun and Lang, J. C. and Islam, Z. and Srajer, G. and Littlewood, P. B. and McWhan, D. B. and Rosenbaum, T. F.},
	month = may,
	year = {2009},
	pages = {405--409},
}

@article{gen_rhombic_2023,
  title = {Rhombic skyrmion lattice coupled with orthorhombic structural distortion in ${\mathrm{EuAl}}_{4}$},
  author = {Gen, Masaki and Takagi, Rina and Watanabe, Yoshito and Kitou, Shunsuke and Sagayama, Hajime and Matsuyama, Naofumi and Kohama, Yoshimitsu and Ikeda, Akihiko and \ifmmode \bar{O}\else \={O}\fi{}nuki, Yoshichika and Kurumaji, Takashi and Arima, Taka-hisa and Seki, Shinichiro},
  journal = {Phys. Rev. B},
  volume = {107},
  issue = {2},
  pages = {L020410},
  numpages = {8},
  year = {2023},
  month = {Jan},
  publisher = {American Physical Society},
  doi = {10.1103/PhysRevB.107.L020410},
  url = {https://link.aps.org/doi/10.1103/PhysRevB.107.L020410}
}

@article{yoshizawa_stripe_2000,
  title = {Stripe order at low temperatures in ${\mathrm{La}}_{2\ensuremath{-}x}{\mathrm{Sr}}_{x}{\mathrm{NiO}}_{4}$ with $0.289\ensuremath{\lesssim}x\ensuremath{\lesssim}0.5$},
  author = {Yoshizawa, H. and Kakeshita, T. and Kajimoto, R. and Tanabe, T. and Katsufuji, T. and Tokura, Y.},
  journal = {Phys. Rev. B},
  volume = {61},
  issue = {2},
  pages = {R854(R)--R857(R)},
  numpages = {0},
  year = {2000},
  month = {Jan},
  publisher = {American Physical Society},
  doi = {10.1103/PhysRevB.61.R854},
  url = {https://link.aps.org/doi/10.1103/PhysRevB.61.R854}
}
\end{document}


\maketitle

\vspace{-2em}
\noindent $^{\dagger}$Corresponding author: Qisi Wang, Email: \href{mailto:qwang@cuhk.edu.hk}{qwang@cuhk.edu.hk}

\vspace{2em}
\noindent \textbf{This PDF file includes:}
\begin{itemize}
    \setlength\itemsep{0em}
    \item[] Supplementary Text
    \item[] Figs. S1 to S3
\end{itemize}

\vspace{1cm}

\newpage

\section*{Supplementary Text}

\underline{Supplementary Note 1: Strain Calibration}\\
\noindent To quantify the applied uniaxial strain on the EuAl$_4$ samples, we determined the shift of structural Bragg peaks using X-ray diffraction. The scattering angle, $2\theta$, for each reflection was extracted from two-dimensional detector images by converting the radial distance $R$ of the peak from the beam center using the relation $\tan(2\theta) = R/D$, where $D$ is the sample-to-detector distance. The interplanar spacing $d$ was subsequently calculated using Bragg's law, $\lambda = 2d \sin(\theta)$, with the X-ray wavelength $\lambda$. This procedure was first performed on an unstrained sample across multiple reciprocal space indices along the $k$ and $\ell$ directions. As shown in Fig. S1, $2\sin(\theta)$ was plotted as a function of the corresponding Bragg indexes. The measurement was then repeated for the sample under applied uniaxial pressure. The relative strain $\varepsilon$ was calculated by comparing the interplanar spacings of the strained state ($d$) to the unstrained reference state ($d_0$) using the relation $\varepsilon = \frac{d - d_0}{d_0}$, providing a direct measure of the lattice deformation along the targeted crystallographic axis.

\begin{figure}[h!]
    \centering
    \includegraphics[width=0.9\linewidth]{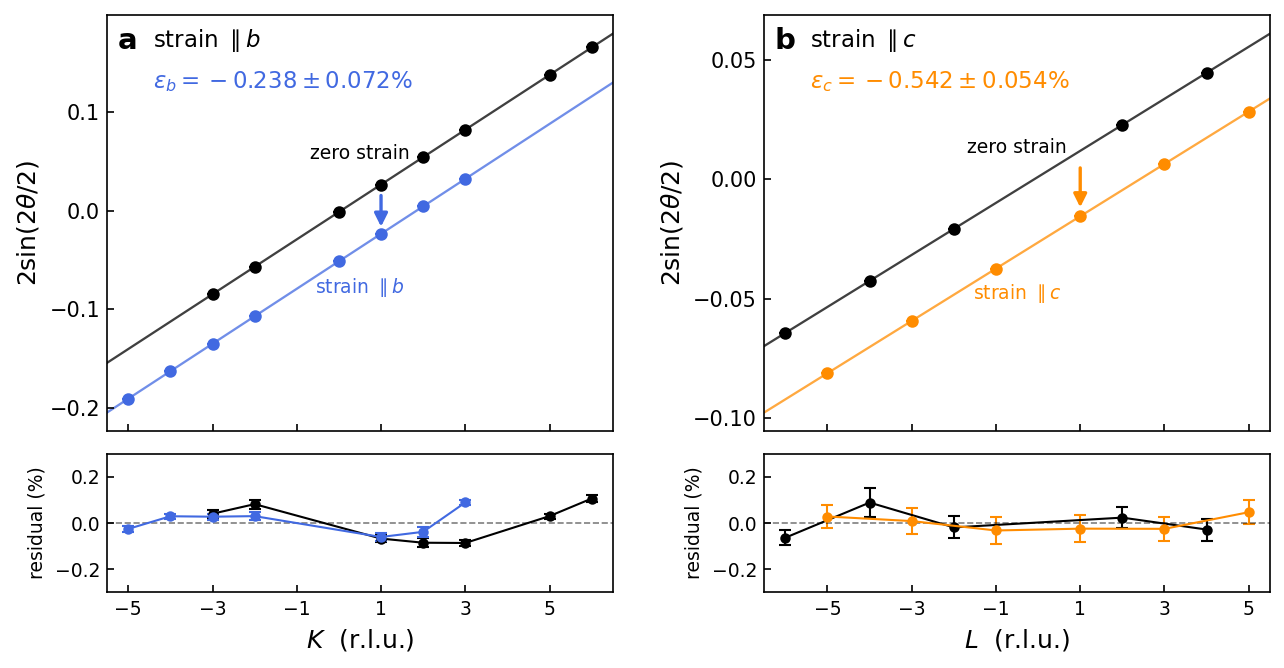}
    \caption{Strain calibration of EuAl$_4$. (a,b) $2\sin(2\theta / 2)$ versus Bragg index measured for the unstrained sample (black) and the sample under uniaxial pressure along $b$- (blue) and $c$- (orange) axis directions. The data of the strained sample is shifted down for better visual presentation. The solid lines are linear least-square fits. Lower panel shows the residuals in percentage.} 
    \label{fig:strain_calibration}
\end{figure}

\noindent\underline{Supplementary Note 2: Onset Temperature of the CDW Phase}\\
\noindent We also investigated the effect of uniaxial strain on the onset temperature ($T_{\text{CDW}}$) of the charge density wave (CDW) phase. Figure S2 tracks the temperature evolution of a representative CDW satellite peak across the transition under various strain conditions. In the unstrained crystal, the CDW satellite emerges via a sharp, well-defined transition between 132~K and 134~K. Applying uniaxial pressure along the $c$-axis suppresses the structural transition, lowering $T_{\text{CDW}}$ to between 120~K and 125~K. Notably, the phase transition becomes significantly broadened under this strain, exhibiting diffuse scattering that persists up to 134~K. Similarly, compressive strain along the $b$-axis slightly reduces $T_{\text{CDW}}$ to between 130~K and 132~K while inducing peak broadening that remains visible up to at least 136~K. Overall, these observations demonstrate that the application of uniaxial strain not only lowers the critical temperature of the CDW but also introduces spatial fluctuations or structural domain fragmentation, in contrast with the abrupt, homogeneous transition observed in the zero-strain state.

\begin{figure}[h!]
    \centering
    \includegraphics[width=\linewidth]{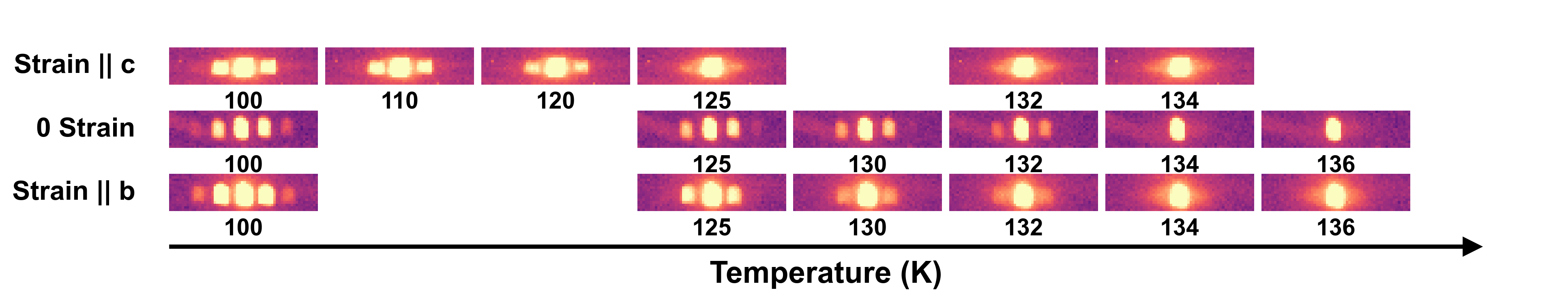}
    \caption{Temperature dependence of the diffraction satellite peaks under varying uniaxial strain. The isolated ROIs span a temperature range of 100~K to 136~K, comparing measurements taken with strain applied along the $c$-axis (top row), no applied strain (middle row), and strain applied along the $b$-axis (bottom row).}
    \label{fig:cdw_onset}
\end{figure}

\begin{figure}[h!]
    \centering
    \includegraphics[width=\linewidth]{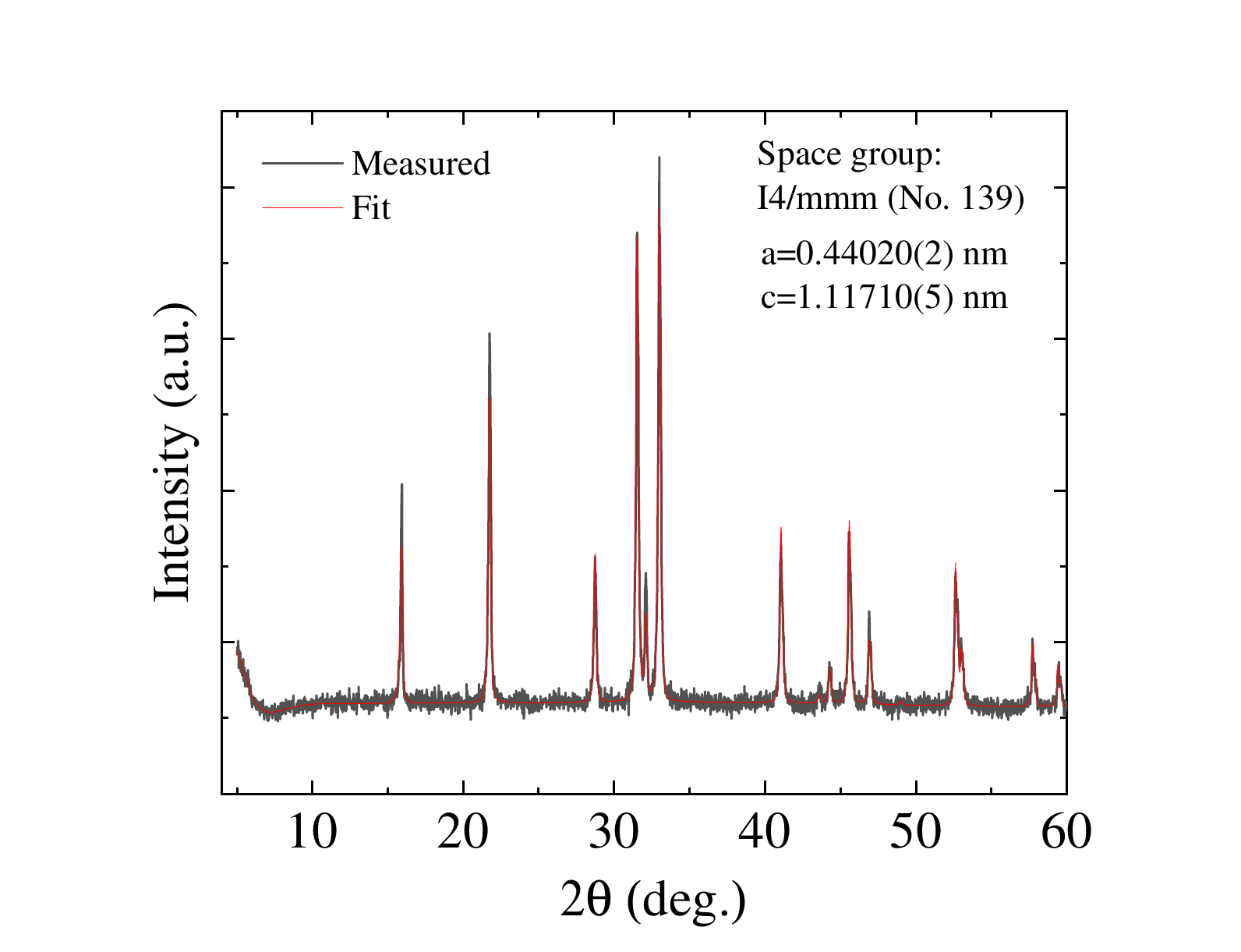}
    \caption{Powder X-ray diffraction data were collected at room temperature using Cu K$\alpha$ radiation on a crushed single crystal. Rietveld refinement was performed using the Profex graphical interface with the BGMN refinement engine \cite{doebelin2015}. No traces of secondary phases were detected.}
    \label{fig:cdw_onset}
\end{figure}